\documentclass[aps,showpacs,twocolumn,superscriptaddress]{revtex4}

\usepackage{amsmath}
\usepackage{amssymb}
\usepackage{amscd}
\usepackage{wasysym}
\usepackage[ansinew]{inputenc}
\usepackage[T1]{fontenc}
\usepackage{ae,aecompl}

\usepackage[rightcaption]{sidecap}

\usepackage[dvips]{graphicx}
\DeclareGraphicsExtensions{.eps}

\newcommand{\be}{\begin{equation}}
\newcommand{\ee}{\end{equation}}
\newcommand{\bea}{\begin{eqnarray}}
\newcommand{\eea}{\end{eqnarray}}
\newcommand{\nn}{\nonumber}
\newcommand{\ket}[1]{|#1\rangle}
\newcommand{\bra}[1]{\left\langle#1\right|}

\newcommand{\eye}{\mbox{$\mbox{1}\!\mbox{l}\;$}}

\newcommand{\LL}{\mathcal{L}}

\renewcommand{\Im}{{\rm Im}}
\newcommand{\tr}{{\rm tr}}

\usepackage{color}
\definecolor{red}{rgb}{1,0,0}

\begin{document}
\bibliographystyle{apsrev}

\title{Decay and fragmentation in an open Bose-Hubbard chain}

\author{G. Kordas}
\affiliation{Institut f\"ur theoretische Physik and Center for Quantum
Dynamics, Universit\"at Heidelberg, 69120 Heidelberg, Germany}
\affiliation{University of Athens, Physics Department, Nuclear \& Particle Physics Section
Panepistimiopolis, Ilissia 15771 Athens, Greece}

\author{S. Wimberger}
\affiliation{Institut f\"ur theoretische Physik and Center for Quantum
Dynamics, Universit\"at Heidelberg, 69120 Heidelberg, Germany}

\author{D. Witthaut}
\affiliation{Max-Planck-Institute for Dynamics and Self-Organization (MPI DS),
37073 G\"ottingen, Germany}

\date{\today }

\begin{abstract}
We analyze the decay of ultracold atoms from an optical lattice
with loss form a single lattice site. If the initial state is dynamically 
stable a suitable amount of dissipation can \emph{stabilize} a 
Bose-Einstein condensate, such that it remains coherent even 
in the presence of strong interactions. 
A transition between two different dynamical phases is observed
if the initial state is dynamically unstable. This transition is 
analyzed here in detail. For strong interactions, 
the system relaxes to an entangled quantum state with remarkable statistical 
properties: The atoms bunch in a few ``breathers'' forming at random 
positions. Breathers at different positions are \emph{coherent},
such that they can be used in precision quantum interferometry
and other applications.
\end{abstract}

\pacs{03.75.Lm, 03.65.Yz, 03.75.Gg}
\maketitle

\section{Introduction}
\label{sec:intro}

Decoherence and dissipation, caused by the irreversible coupling of a 
quantum system to its environment, represent a major obstacle for a 
long-time coherent control of quantum states.
Sophisticated methods have been developed to maintain coherence 
also in the presence of dissipation with applications in quantum control
and quantum information processing \cite{Niel00,Gard04}.
Only recently a new paradigm has been put forward: 
Dissipation can be used as a powerful tool to steer the dynamics
of complex quantum systems if it can be accurately controlled.
It was shown theoretically that dissipative processes can be constructed 
which allows one to prepare pure states for quantum computation 
\cite{Diel08,Diel11}, to implement universal quantum computation
\cite{Vers09} or to deterministically generate entangled quantum
states \cite{Kast11,Krau10}.

Ultracold atoms in optical lattices provide a distinguished system
to realize new methods of quantum control and quantum state
engineering \cite{Bloc08}. Both the coherent dynamics of the atoms 
as well as dissipative processes can be accurately controlled, 
including the localized  manipulation with single-site resolution 
\cite{06nlres,Geri08,Wurt09,Bakr09,Sher10,Gros10,Baron13}.
In this article we analyze the dynamics induced by the interplay
of localized particle dissipation and strong atom-atom interactions.
If the interaction strength exceeds a threshold,  two meta-stable 
equilibria emerge which can be used to prepare either an almost
pure Bose-Einstein condensate or a macroscopically entangled
``breather'' state.

The meta-stable breather states show remarkable statistical 
properties: The atoms relax to a coherent superposition of
bunches localized at different lattice positions. 
Driven by particle loss and interactions, almost all atoms localize in 
one of the non-dissipative wells. The meta-stable state corresponds
to a \emph{coherent} superposition of these localized modes
and thus to a macroscopically entangled quantum state. Because
of the tunable large number of atoms forming the breather state,  
they may serve as a distinguished probe of decoherence and the 
emergence of classicality. Furthermore, the breather states 
generalize the so-called NOON states enabling interferometry 
beyond the standard quantum limit \cite{Boll96,Giov04}. 
As particle loss is an elementary and omnipresent dissipation process,
this method may be generalized to a variety of open quantum systems
well beyond the dynamics of ultracold atoms, e.g. to optical
fiber setups~\cite{Reg11} or hybrid quantum systems~\cite{Tom12,Schm12,Wim11}.

The paper is organized as follows.
After introducing the model system in Sec.~\ref{sec:loss}, 
we analyze breather states in small systems, 
which allow for a numerically exact simulation of the quantum 
many-body dynamics in Sec.~\ref{sec:trimer}.
In extended lattices discussed in Sec.~\ref{sec:lattice},  the localized 
modes correspond to so-called
discrete breathers. The emerging meta-stable quantum state is 
more complex, as the atoms can localize in a variety of lattice sites. 
Nevertheless, one can identify ``breather-states'' by the number
fluctuations and the correlations between neighboring sites.
The formation of breather states can be understood to a large 
extent within a semi-classical phase space picture introduced 
in Sec.~\ref{sec:semi}.
We analyze the flow of phase space distribution functions such as the Wigner
or the Husimi function. To leading order it is given by a classical
Liouvillian flow which is equivalent to a dissipative Gross-Pitaevskii
equation. The emergence of breather states can then be linked to 
a classical bifurcation of the associated mean-field dynamics.
While this semiclassical approach obviously cannot describe the
coherence of the quantum state or the formation of entanglement, 
it correctly predicts the critical interaction strength above which
breather states are formed.

\section{Particle loss in an optical lattice}
\label{sec:loss}

Optical lattices offer unique possibilities in controlling the quantum
dynamics of ultracold atoms \cite{Bloc08,Bloc08b}. In particular, experimental 
parameters such as the strength of the atom-atom interactions can 
be readily tuned by a variation of the lattice depth.
Recently, several experiments demonstrated a \emph{local} control
of the atomic dynamics. Single site access can be implemented optically 
either by increasing the lattice period \cite{Albi05,Gros10} or by pushing 
the resolution of the optical imaging system to the limit \cite{Bakr09,Sher10}.
Furthermore, the advanced imaging systems in these experiments
enable a precise 
measurement of the atom number per site. 
An even higher resolution can be realized by a focused electron beam 
\cite{Geri08,Wurt09}. However, 
the interaction of the electron beam with the atomic
cloud is generally dissipative: Atoms are ionized and then removed from the
lattice by a static electric field. At the same time, this methods enables the
detection of single atoms with outstanding spatial resolution.

The coherent dynamics of ultracold atoms in deep optical lattices
is described by the celebrated Bose-Hubbard Hamiltonian \cite{Jaks98}
\bea
  \hat H = - J  \sum \nolimits_{j} \left( \hat a_{j+1}^{\dagger} \hat a_j +
                  \hat a_{j}^{\dagger} \hat a_{j+1} \right)
         + \frac{U}{2} \sum \nolimits_j 
           \hat a_{j}^{\dagger}  \hat a_{j}^{\dagger} 
               \hat a_{j}  \hat a_{j},
    \label{eqn-hami-bh}
\eea
where $\hat a_j$ and $\hat a_j^\dagger$ are the bosonic annihilation and 
creation operators in mode $j$, $J$ denotes the tunneling
matrix element between the wells and $U$ the interaction strength. 
We set $\hbar = 1$, thus measuring energy in frequency units.
This model assumes that the lattice is sufficiently deep, such that
the dynamics takes place in the lowest Bloch band only.
Throughout this paper we consider finite lattices with $M$ sites with 
periodic boundary conditions, i.e. we identify the sites $j=0$ and
$j=M$.

In this article we analyze the non-equilibrium dynamics triggered by 
localized dissipation implemented either by a resonant laser or a 
focused electron beam. The atoms are removed rapidly and irreversibly 
from the lattice, such that the dissipative dynamics can be described by
a Markovian Master equation,
\be 
  \frac{d}{dt} {\hat \rho} =  -i [\hat H,\hat \rho]  + \LL \hat \rho.
  \label{eqn:master}
\ee
Particle loss is described by the Liouvillian 
\cite{Breu02,11leaky1,11leaky2,Kepe12,bar11} 
\be
  \LL_{\rm loss} \hat \rho = -\frac{1}{2} \sum \nolimits_j \gamma_j \left(
     \hat a_{j}^{\dagger} \hat a_{j}  \hat \rho
     + \hat \rho \hat a_{j}^{\dagger} \hat a_{j} 
     - 2 \hat a_j \hat \rho \hat a_{j}^{\dagger}  \right),
\ee
where $\gamma_j$ denotes the loss rate at site $j$.
Furthermore, the atoms experience phase  noise due to collisions with 
the background gas \cite{Angl97,Truj09,pol12} or the absorption and spontaneous emission
of photons from the lattice beams \cite{Pich10}. This dissipation process
is described by the Liouvillian
\be 
  \LL_{\rm noise} \hat \rho = - \frac{\kappa}{2} \sum \nolimits_j
    \left( \hat n_j^2 \hat \rho + \hat \rho \hat n_j^2 
          - 2 \hat n_j \hat \rho \hat n_j  \right).
\ee
Phase noise can be made very small, e.g. by detuning the optical lattice 
far from the atomic resonance such that we can assume $\kappa = 0$ in 
most simulations. A detailed analysis of decoherence due to phase noise
is then provided in Sec.~\ref{sec:ent_dec}.

\begin{figure}[tb]
\centering
\includegraphics[width=4cm, angle=0]{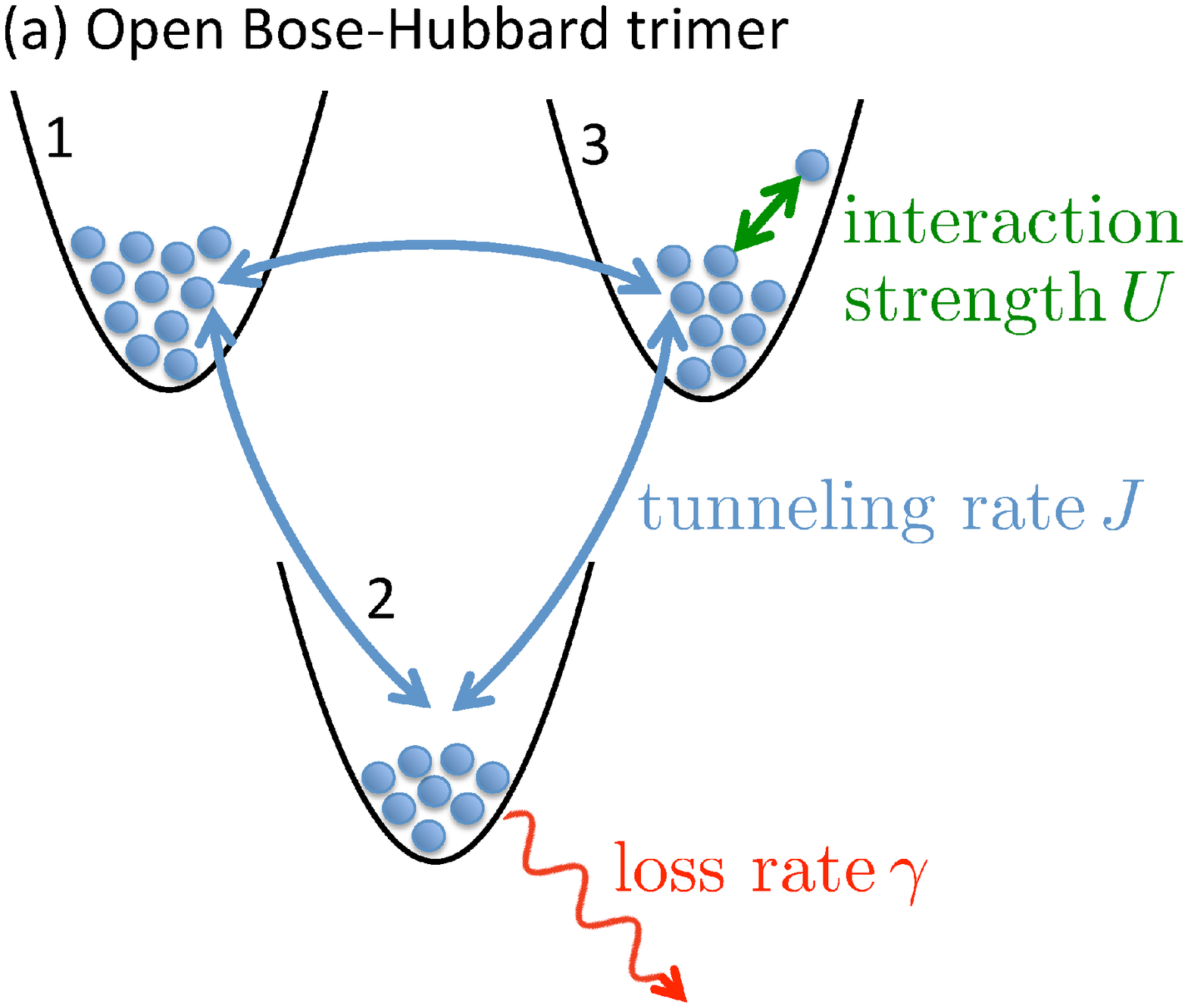}
\includegraphics[width=4cm, angle=0]{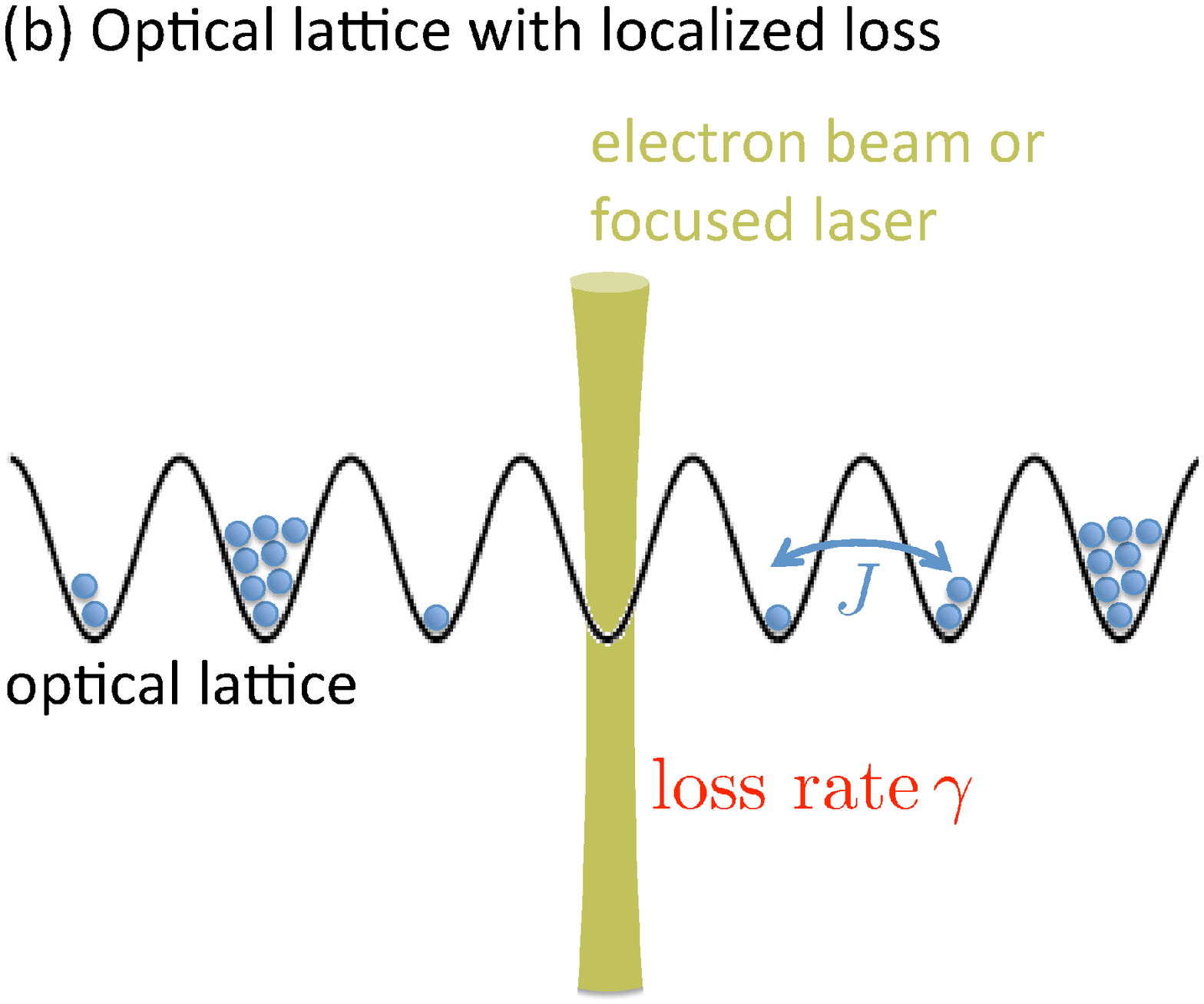}
\caption{\label{fig:setup}
(Color online)
The model systems studied in the present paper:
(a) An open Bose-Hubbard trimer with loss from
site 2 and periodic boundary conditions,
(b) An extended one-dimensional optical lattice
with localized loss from a single lattice site.
}
\end{figure}

For numerical simulations, we will make use of the quantum jump method 
\cite{Dali92,Carm93}, where the density matrix $\hat \rho$ is decomposed into 
state vectors, 
\be
   \hat \rho = \frac{1}{L} \sum_{\ell = 1}^L \ket{\Psi_\ell} \bra{\Psi_\ell},
\ee
whose continuous evolution is interrupted by stochastic quantum jumps.
The continuous evolution is determined by the Schr\"odinger equation with
the effective non-hermitian Hamiltonian
\be
   \hat H_{\rm eff} = \hat H 
      - \frac{i}{2} \sum \nolimits_j \gamma_j \hat a_j^\dagger \hat a_j 
     - \frac{i \kappa}{2}  \sum \nolimits_j \hat n_j^2 \, .
    \label{eqn:Heff}  
\ee
Since $\hat H_{\rm eff}$ is non-hermitian, the state vector 
$\ket{\Psi}$ must be renormalized after every time-step.
In the case of particle loss, the state vector jumps according to
\be
   \ket{\Psi} \rightarrow \frac{\hat a_j  \ket{\Psi}}{\|  \hat a_j  \ket{\Psi} \|}
\ee
with a probability
\be 
   \delta p = \gamma_j \bra{\Psi} \hat a_j^\dagger \hat a_j \ket{\Psi} \delta t 
\ee
during a short time interval $\delta t$. The full density matrix
is recovered by averaging over many of these random trajectories 
in state space.

\begin{SCfigure*}
\centering
\includegraphics[width=12cm, angle=0]{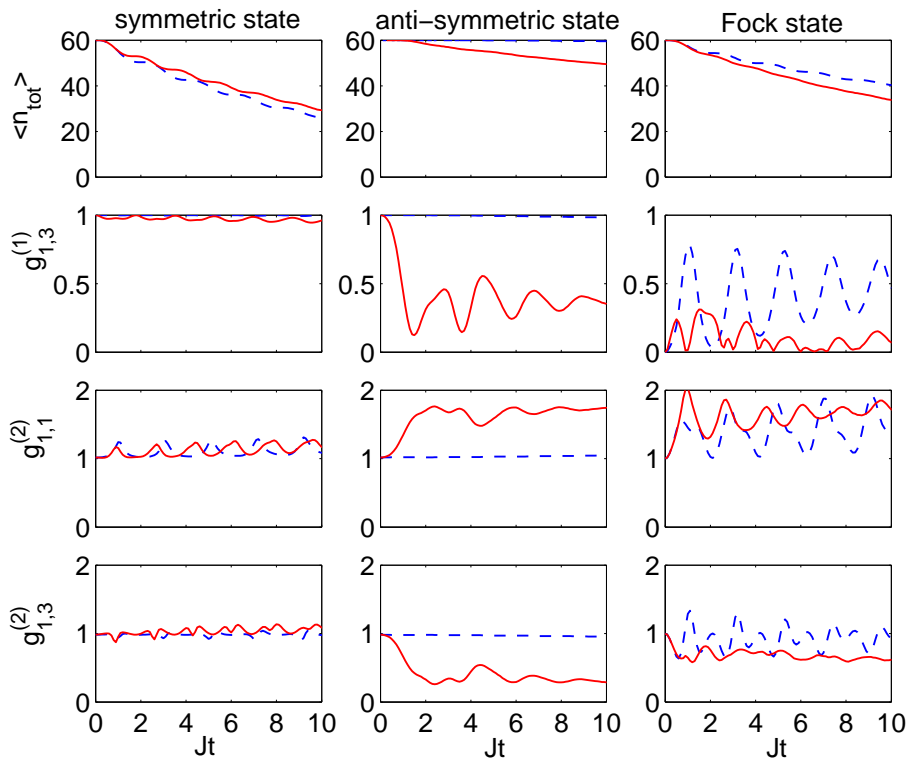}
\caption{\label{fig:dynamics}
(Color online)
Dynamics of the atom number and the correlation functions in an
open Bose-Hubbard trimer with loss from site 2 for weak interactions
($U = 0.01 J$, dashed blue line) and strong interactions 
($U = 0.1 J$, solid red line). Plotted is the total particle
number $n_{\rm tot}$ (first row), the phase coherence between
the sites 1 and 3 $g^{(1)}_{1,3}$ (second row), the number fluctuations $g^{(2)}_{1,1}$
(third row) and the number correlations between
the sites 1 and 3 $g^{(2)}_{1,3}$ (fourth row).
The dynamics has been simulated for three different initial states: A BEC
with symmetric wave function (left), a BEC with an anti-symmetric
wave function (middle) and a Fock state (right).
The loss rate is $\gamma_2 = 0.2 J$ and the initial populations are $n_1(0)=n_3(0)=30$, $n_2(0)=0$ in all cases.
Time is given as $Jt$ in units of the tunneling time.
}
\end{SCfigure*}

\section{Decay in an open triple-well trap}
\label{sec:trimer}

To begin with, we consider the open Bose-Hubbard trimer as an
elementary model system, allowing for numerical exact solutions for rather large particle numbers. 
Still, this model already exhibits the different dynamical phases
we aim to understand. A sketch of this system is provided in
Fig.\ref{fig:setup} (a). The bosons tunnel at a rate $J$ between three 
lattice sites with periodic boundary conditions. Loss occurs only 
from the site $j=2$ at a rate $\gamma_2$. The system is 
mirror-symmetric with respect to an exchange of sites 1 and 3. 

\subsection{Atomic correlations}

The most obvious effect of particle dissipation is the decrease of
the total particle number $n_{\rm tot}$ in the lattice, which is
shown in the top row of Fig.~\ref{fig:dynamics}. We simulate
the dynamics for three different initial states for weak ($U=0.01J$)
and strong ($U=0.1J$) interactions. A pure Bose-Einstein
condensate with an (anti-) symmetric wavefunction
\be
   \ket{\Psi_\pm} = \frac{1}{2^N \, \sqrt{N!}} 
       (\hat a_1^\dagger \pm \hat a_3^\dagger  )^N \ket{0}
\ee
and the Fock state 
\be
   \ket{\Psi_F} = \frac{1}{2 \sqrt{(N/2)!} } 
         (\hat a_1^\dagger)^{(N/2)}  (\hat a_3^\dagger)^{(N/2)} \ket{0},
\ee 
assuming that the initial particle number $N$ is even.
The most interesting observation here is that the decay is very slow 
for the anti-symmetric initial state $\ket{\Psi_-}$. Indeed, this state
is a stationary state of the master equation (\ref{eqn:master}) for $U=0$,
such that decay is absent in the non-interacting limit
(cf.~\cite{Kepe12}).
The physical reason for this is the destructive interference
of atoms tunneling from sites 1 and 3 to the leaky site 2.
In the case of strong interactions, tunneling is allowed but weak.
Localized states, which will be refered to as breather states, 
form at the non-dissipative sites. The formation and properties 
of these states is analyzed in detail in the present paper.

The dissipative dynamics drives the atoms to a very different quantum
state  depending on the initial state and the interaction strength $U$.
To characterize these states we analyze the first and second order 
correlation functions between different sites of the lattice.
The coherence of the many-body quantum state is characterized by
the first-order correlation function between the wells $j$ and $\ell$,
\be
  g^{(1)}_{j,\ell} = \frac{\langle  \hat a_j^\dagger \hat a_\ell \rangle }{
                             \sqrt{ \langle \hat n_j \rangle \langle \hat n_\ell \rangle}},
\ee
which is plotted in the second row of Fig.~\ref{fig:dynamics}.
The symmetric initial state  $\ket{\Psi_+}$ is stable for all values of
the interaction strength $U$ and the BEC remains approximately
pure. In this case, particle dissipation can even increase the purity
and coherence of the condensate. This counter-intuitive feature 
was discussed in detail in \cite{08stores,09srlong,11leaky1,11leaky2}.
The anti-symmetric state $\ket{\Psi_-}$ is stable only if interactions
are weak. For $U=0.1 J$ one observes a sharp decrease of first-order 
correlation which indicates the destruction of the condensate. The initial state
is dynamically unstable such that the atoms relax to a different 
meta-stable equilibrium state, the breather state.

Density fluctuations and correlations
are characterized by the second order correlation function
\be
  g^{(2)}_{j,\ell} = \frac{ \langle  \hat n_j \hat n_\ell \rangle }{
                             \langle \hat n_j \rangle \langle \hat n_\ell \rangle}.
\ee
For $j = \ell$, this expression reduces to the normalized 
second moment of the number operator
$\langle  \hat n_j ^2\rangle/\langle  \hat n_j \rangle^2$,
which quantifies the number fluctuations in the $j$th well. 
The evolution of the number fluctuations  and correlations
are shown in Fig.~\ref{fig:dynamics} in the bottom panels.
While these quantities are essentially constant for a BEC
with a symmetric wave function $\ket{\Psi_+}$, strong 
anti-correlations develop for the initial state $\ket{\Psi_-}$
in the regime of strong interactions.
The (anti-) correlations are also found for the Fock state $\ket{\Psi_F}$,
whose experimental preparation can be significantly easier.
These results show that the atoms bunch at one of the
non-dissipative lattice sites, while the other sites are 
essentially empty. Nevertheless, as we are going to discuss in detail in section \ref{IIIC}, the two contributions
localized either at site 1 or 3 remain \emph{coherent}.
The atoms thus relax deterministically to 
a macroscopically entangled
state, also called a Schr\"odinger cat state 
(cf.~\cite{Leib05}).
We will refer to these states as ``breather'' states as they 
correspond to the so-called discrete breathers in extended
lattices in the semiclassical limit \cite{Camp04,Flac08,Ng09}. 
This correspondence will
be discussed in detail in Sec.~\ref{sec:semi}. 

\begin{figure}[tb]
\centering
\includegraphics[width=8.5cm, angle=0]{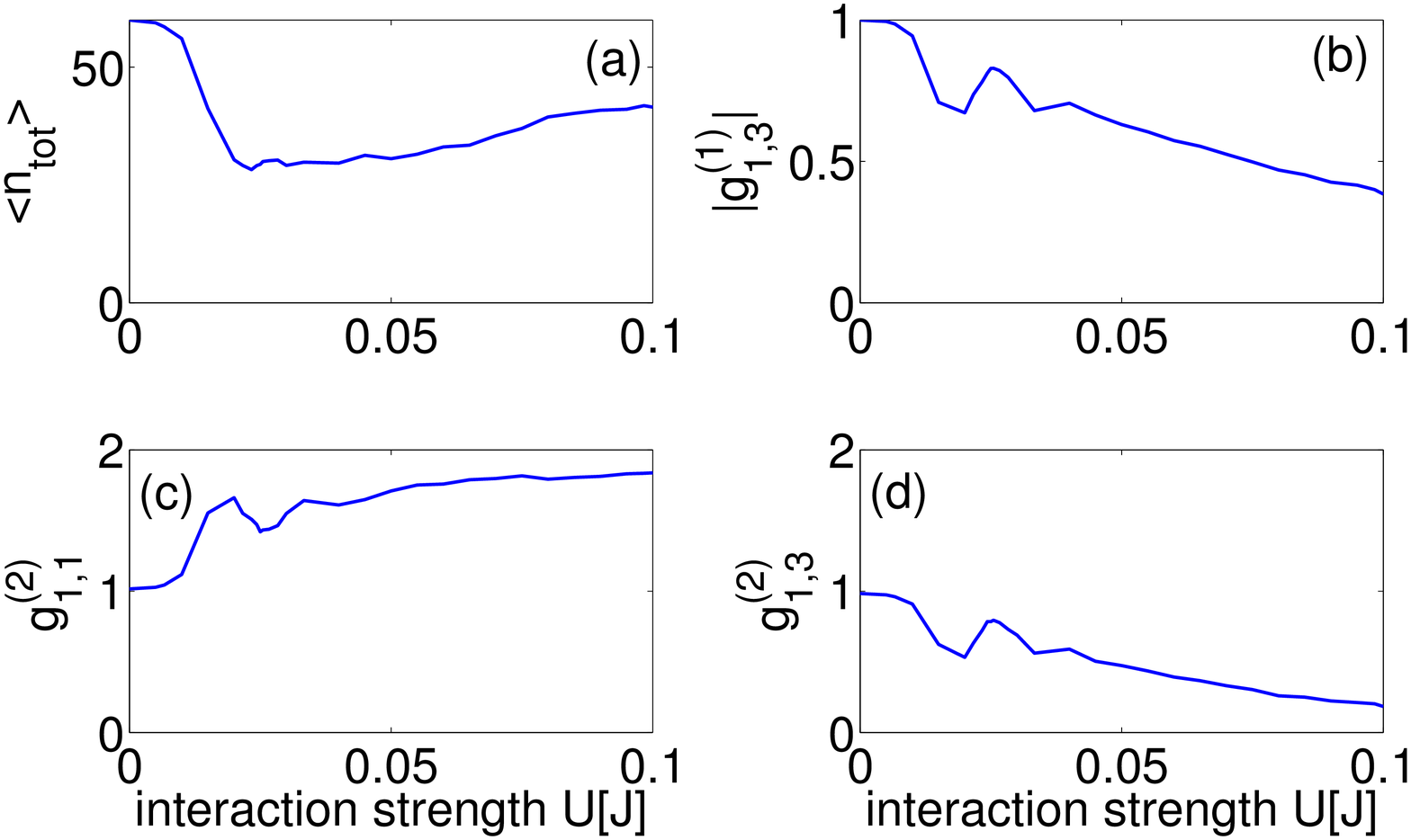}
\caption{\label{fig:trimer_qjU}
(Color online)
Onset of breather formation in a triple-well trap for strong atomic 
interactions.
Shown are (a) the total particle number 
$\langle \hat n_{\rm tot}\rangle$, (b) the phase coherence 
$g^{(1)}_{1,3}$ and (c,d) the density correlation functions $g^{(2)}_{1,1}$ and $g^{(2)}_{1,3}$
as a function of the interaction strength $U$ after a fixed propagation time 
$t_{\rm final} = 50 J^{-1}$ for $\gamma_2=0.2 J^{-1}$.
The initial state is a pure BEC in the state $\ket{\Psi_-}$ 
with $N=60$ atoms.
}
\end{figure}

\subsection{Transition to the breather regime}
 
The meta-stable breather states exists only for strong atomic interactions.
The onset of breather formation is analyzed in Fig.~\ref{fig:trimer_qjU}, 
where we have plotted the total particle number as well as the first 
and second order correlation after a fixed propagation time 
$t_{\rm final}=50 \, \rm s$ as a function of the interaction strength $U$.
As one can see for weak interactions, $U \lesssim 0.01\,\rm s^{-1}$, 
the BEC remains almost pure and the density-density correlation 
function are approximately equal to unity. 
The characteristic properties of a breather state, strong number fluctuations
and anti-correlations between neighboring sites, are observed only for
$U \apprge 0.01\,\rm s^{-1}$. The transition to the breather regime can 
be understood within a semiclassical phase space picture which will be 
discussed in detail in Sec.~\ref{sec:semi}. This approach predicts a bifurcation
of meta-stable states at a critical interaction strength 
$Un_{\rm tot} = 0.4 J$. Before we
come back  to this issue, we first  characterize the quantum properties of 
the breather states in more detail.

\subsection{Characterization and Interferometry of the breather state}
\label{IIIC}

\begin{figure}[tb]
\centering
\includegraphics[width=8cm, angle=0]{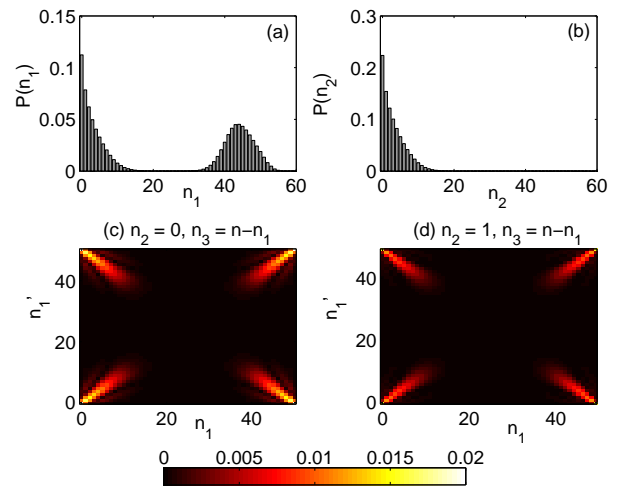}
\caption{\label{fig:fcs}
(Color online)
(a,b)
Full counting statistics of the breather state at time $t  = 10 \, J^{-1}$
in the first and second well.
(c,d)
Density matrix $\rho(n_1,n_2,n_3; n_1', n_2', n_3')$ of a breather
state at time $t  = 10 \, J^{-1}$ for a fixed particle number $n = 50$ 
and $n_2 = n_2'= 0$ (c) and $n_2 = n_2' = 1$ (d), respectively.
Parameters are $U = 0.1 \, J$ and $\gamma_2 = 0.2 \, J$.
The initial state is a pure BEC with $N=60$
atoms and an anti-symmetric wave function (\ref{eqn_initial-as}). 
}
\end{figure}

In a breather state a large number of atoms localize at a single 
lattice site, leaving the neighboring sites essentially empty.
To make this statement more precise, we analyze the full counting 
statistics and the coherence of the many-body quantum state in
detail.
Figure \ref{fig:fcs} (a,b) shows the full counting statistics of the atom 
number in well 1 and 2, respectively, at time $t= 10 \, J^{-1}$ after a 
breather state has formed. 
The most important result is that the probability distribution 
$P(n_1)$ becomes bimodal: Either a breather forms in the 
first well ($n_1$ large) or in the third well ($n_1$ almost zero). 
The second well is almost empty for large values of the interaction 
constant $U$.  This stabilizes the breather state as only few atoms 
are subject to particle loss.
For intermediate values of the interaction constant $U$, one
also finds the characteristic bimodal number distribution in the
first well. However, the atom number in the second well 
is larger, such that decay is much stronger.

The two breathers in site 1 and 3 are fully coherent, even
for large interactions. To analyze the coherence of the the 
many-body quantum state $\hat \rho(t)$ in more detail,
we first note that $\hat \rho(t)$ can be written as the incoherent 
sum of contributions with different total particle number $n$:
\be
   \hat \rho(t) = \sum \nolimits_n  p_n(t) \,  \hat \rho^{(n)}(t).
\ee
There are no coherences between the contributions 
$\hat \rho^{(n)}(t)$ as particle loss proceeds via incoherent 
jumps only. 
Numerical results for the density matrix $\hat \rho^{(n)}(t)$
with $n=50$
are shown in Fig.~\ref{fig:fcs} (c,d) at time $t = 10 \, J^{-1}$ 
after the formation of a breather state. We have plotted the 
matrix elements of $\hat \rho^{(n)}(t)$ for a subset of matrix 
indices, fixing $n_2 = n_2' = 0$ or $n_2 = n_2' = 1$, respectively. 
For this plot we simulated the dynamics with the quantum jump 
method using $L=3000$ stochastic trajectories in total. 
One observes that the coherences, i.e. the off-diagonal matrix
elements of the projected density matrix assume their maximum
possible values,
\be
   |\rho_{n_1,n'_1}|^2 \approx | \rho_{n_1,n_1}| \, |\rho_{n'_1,n'_1}|. 
\ee
This shows that the two breathers formed at lattice sites 1 and 3 
are indeed fully \textit{coherent}. 
Breather states with different total particle number are generally 
not coherent as discussed above. However, this neither affects
the entanglement of the atoms nor its use in quantum interferometry.

\begin{figure}[tb]
\centering
\includegraphics[width=8cm, angle=0]{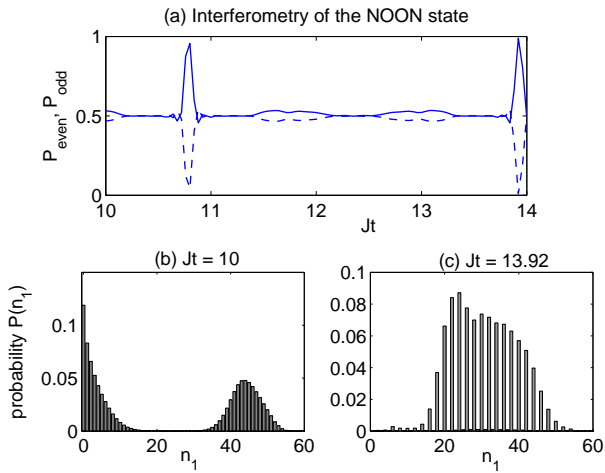}
\caption{\label{fig:interfere}
(Color online)
Interferometry of the NOON state according to the time evolution
$\hat U(t) = \exp[-J(\hat a_1^\dagger \hat a_3 - \hat a_3^\dagger \hat a_1)t]$.
(a) Probability to detect an even (solid line) and an odd number of
atoms (dashed line) at site 1 as a function of time.
Lower panels: Full counting statistics at site 1 at (b) the beginning of the 
interferometer stage $t = 10 J^{-1}$ and (c) during the interferometer stage 
at $t = 13.92 J^{-1}$, where $P_{\rm even} = 1$.
The breather state is generated starting from a BEC with an anti-symmetric
wave function as shown in Fig.~\ref{fig:dynamics} with $U = 0.1 J$ and
$\gamma_2 = 0.2 J$.
During the interferometer stage we assume that $U = \gamma_2 = 0$.
}
\end{figure}

Due to the almost perfect coherence of the modes, breather states
enable new applications in precision quantum metrology. In particular, 
they generalize the so-called NOON states $\ket{n,0,0} + e^{i \alpha} \ket{0,0,n}$ 
which enable precision interferometry beyond the standard quantum limit 
\cite{Giov04}. 
Breather states can be written as a superposition of states of the form 
\be
     \ket{n_1,n_2,n-n_1-n_{2}} + e^{i \alpha} \ket{n-n_1-n_{2},n_2,n_1}. 
\ee
That is, the coherence of wells $1$ and $3$ is guaranteed as in an
ordinary NOON state but the total number of atoms forming the 
NOON state varies statistically. Nevertheless, this is sufficient for
precision interferometry. 

Starting from the breather state analyzed in the preceding section,
we consider an interferometric measurement, where the modes 
(lattice sites) $1$ and $3$ are mixed. Assuming that interactions
(by tununig a Feshbach resonance)
and losses are switched off, the dynamics during the interferometer 
stage is given by the time evolution operator
\be
   \hat U_{\rm interferometer} = \exp[- i\hat{H}_{\rm mix}t ],
\ee   
where $\hat{H}_{\rm mix}=iJ(\hat{a}_1^\dagger \hat{a}_3 - \hat{a}_3^\dagger \hat{a}_1)$.
In analogy to the parity observable in NOON state interferometry 
\cite{Boll96},
we record the probability to detect either an even or an odd number
of atoms in site 1. Such a measurement is automatically realized
by the optical imaging apparatus in the experiments \cite{Bakr09,Sher10}.

This probability $P_{\rm even, odd}$ to detect an even or an
odd number of atoms is plotted in Fig.~\ref{fig:interfere} (a) as a 
function of time. $P_{\rm even}$ approaches unity periodically
at times
\be
   t_{\rm rev} = \left(n+ \frac{1}{4} \right) \pi \, J^{-1} ,
   \qquad n = 0,1,2,\ldots,
  \label{eqn:trev}
\ee
which unambiguously proves the coherence of the breather
state. Figure \ref{fig:interfere} (b,c) shows the full counting statistics
in site 1 at the beginning of the interferometer stage at $t = 10 J^{-1}$  
and during the interferometer stage at $t = 13.92 J^{-1}$. Destructive
interference forbids to detect an odd number of atoms at this time,
such that $P_{\rm even}(t)$ approaches unity. 

The interference fringes observed for $P_{\rm even,odd}(t)$
are extremely sharp, which enables
precision measurement beyond the standard quantum limit.
In the present setup, the detection of a fringe reveals the value 
of the tunneling rate $J$ with ultra-high precision via equation
(\ref{eqn:trev}). Different quantities can be measured by  
a modified interferometry scheme as described in \cite{Boll96}.
An important but very difficult goal is to increase the number 
of atoms forming a NOON state (see, e.g., \cite{Afek10}),
as the measurement uncertainty of this method scales inversely
with the particle number $N$.
This goal may be archived with the breather states discussed
here which are readily generated also for large samples.

\subsection{Entanglement and decoherence}
\label{sec:ent_dec}
The atoms in a breather or NOON state are strongly entangled:
If some atoms are measured at one site, then the remaining 
atoms will be projected onto the same site with overwhelming probability.
To unambiguously detect this form of multi-partite entanglement,
we analyze the variance of the population imbalance 
$\Delta(\hat n_3 - \hat n_1)^2$, which scales as $\sim n_{\rm tot}^2$
for a breather state, while it is bounded by $n_{\rm tot}$ for a pure 
product state, $n_{\rm tot}$ being the total atom number.
The variance can thus serve as an entanglement criterion, 
if the quantum state is pure or, more importantly, if one can
assure that a large value of the variance is not due to an incoherent 
mixture of states localized at site 1 or 3.

\begin{figure}[tb]
\centering
\includegraphics[width=8cm, angle=0]{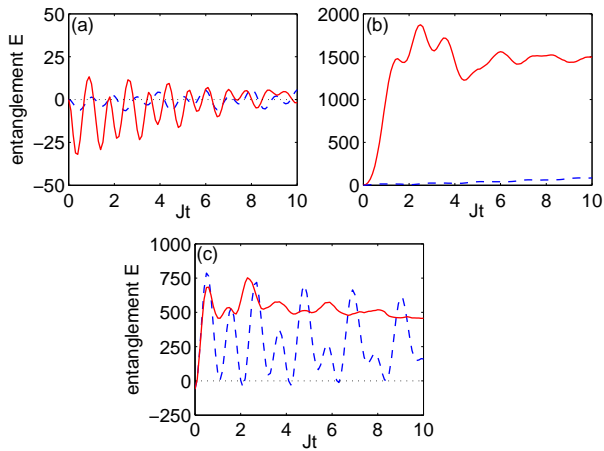}
\caption{\label{fig:entanglement}
(Color online)
Evolution of the entanglement parameter 
(\ref{eqn:ent_para}) for three different
initial states: 
(a) a BEC with symmetric wave function $\ket{\Psi_+}$, 
(b) a BEC with an anti-symmetric wave function  $\ket{\Psi_-}$ and 
(c) a Fock state $\ket{\Psi_F}$ .
Parameters are $\gamma_2 = 0.2 J$,
$U = 0.01 J$ (dashed blue lines) and 
$U = 0.1 J$ (solid red line), respectively.
}
\end{figure}

We assume that a quantum state is decomposed into pure states,
$\hat \rho = L^{-1} \sum_{a=1}^L \ket{\psi_a}\bra{\psi_a}$,
as it is automatically the case in a quantum jump simulations 
\cite{Dali92}.
We then introduce the entanglement parameter
\bea
         \label{eqn:ent_para}    
    E_{r,q} &:=& \langle (\hat n_r - \hat n_q)^2 \rangle
     - \langle \hat n_r - \hat n_q \rangle^2 
     - \langle \hat n_r + \hat n_q \rangle  \\
    &&  - \frac{1}{2L^2} \sum_{a,b} 
     \left[ \langle (\hat n_r - \hat n_q) \rangle_a 
         - \langle (\hat n_r - \hat n_q) \rangle_b \right]^2, \nn
\eea
for the wells $(r,q)$, where $\langle \cdot \rangle_{a,b}$ denotes 
the expectation value in the pure state $\ket{\psi_{a,b}}$.  The last 
term in the parameter $E_{r,q}$ corrects for the possibility of an 
incoherent superposition of states localized at sites $1$ and $3$. 
For a separable quantum state one can now show that 
$E_{j,k}<0$ such that a value $E_{j,k}>0$ unambiguously proves
entanglement  of the atoms. 
The detailed derivation is given in appendix \ref{sec:ent}. 

Figure \ref{fig:entanglement} shows the evolution of the 
entanglement parameter $E_{1,3}(t)$ for three different 
initial states. The symmetric state $\ket{\Psi_+}$ remains
close to a pure BEC, such that $E_{1,3}(t) \approx 0$ for
all times.
In contrast, the anti-symmetric state $\ket{\Psi_-}$ and the
Fock state $\ket{\Psi_F}$ relax to strongly entangled breather 
states if interactions are sufficiently strong. 
In this case we observe large positive values of the entanglement 
parameter $E_{1,3}(t) \approx 1500$ and $E_{1,3}(t) \approx 500$,
respectively, which clearly reveals the presence of many-particle 
entanglement.
Notably, entanglement is also generated for the Fock state 
$\ket{\Psi_F}$ in the regime of weak interactions $U=0.01 J$.
However, this is only a transient phenomena caused by interference 
effects. The breather states formed in the case of strong interactions
are metastable such that the generated entanglement persists for long times
until all atoms decay from the trap. Thus, localized particle dissipation
enables the robust, deterministic generation of entanglement only in the
presence of strong interactions.

\begin{figure}[tb]
\centering
\includegraphics[width=8cm, angle=0]{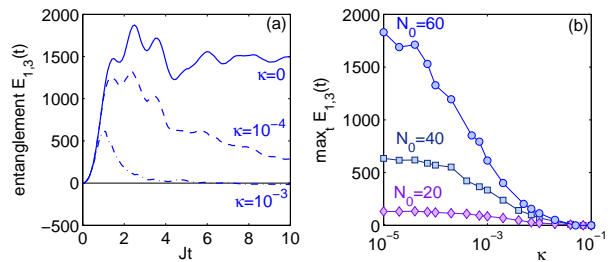}
\caption{\label{fig:ent}
(Color online)
Entanglement and decoherence of a breather state in the presence
of phase noise
(a) Evolution of the entanglement parameter (\ref{eqn:ent_para})
for the anti-symmetric initial state $\ket{\Psi_-}$ and
$\kappa=0$ (solid line), $\kappa = 10^{-4}J$ (dashed line)
and $\kappa=10^{-3}J$ (dash-dotted line) and $N(0)=60$.
(b) Temporal maximum of the entanglement parameter ${\rm max}_t E_{1,3}(t)$
as a function of the phase noise rate $\kappa$ for the anti-symmetric
initial state $\ket{\Psi_-}$ and different particle numbers.
Parameters are $U=0.1 J$ and $\gamma_2  = 0.2 J$.
}
\end{figure}

Furthermore, entangled breather states provide a sensitive probe 
for environmentally induced decoherence. Figure \ref{fig:ent}  (a)
shows the evolution of the entanglement parameter $E_{1,3}(t)$
for three different values of the strength of phase noise $\kappa$
starting from the anti-symmetric initial state $\ket{\Psi_-}$. 
Entanglement is generated in all cases, but $E_{1,3}(t)$ rapidly
decreases again  when $\kappa$ is large due to the decoherence 
of the breathers. 
Notably, one finds  strong number fluctuations $g^{(2)}_{1,1} > 1$
and anti-correlations  $g^{(2)}_{1,3} < 1$ also in the presence of 
strong phase noise, but interferometry is no longer possible.
Figure \ref{fig:ent}  (b)
shows the maximum value of $E_{1,3}(t)$ realized in the presence of 
phase noise. Entanglement decreases with the noise rate $\kappa$,
in which breather states with large particle numbers are most sensitive.
However, entanglement persists up to relatively large values of 
$\kappa \approx 10^{-2}J$ in all cases.

\section{Semiclassical interpretation}
\label{sec:semi}

\begin{figure*}[tb]
\centering
\includegraphics[width=16cm, angle=0]{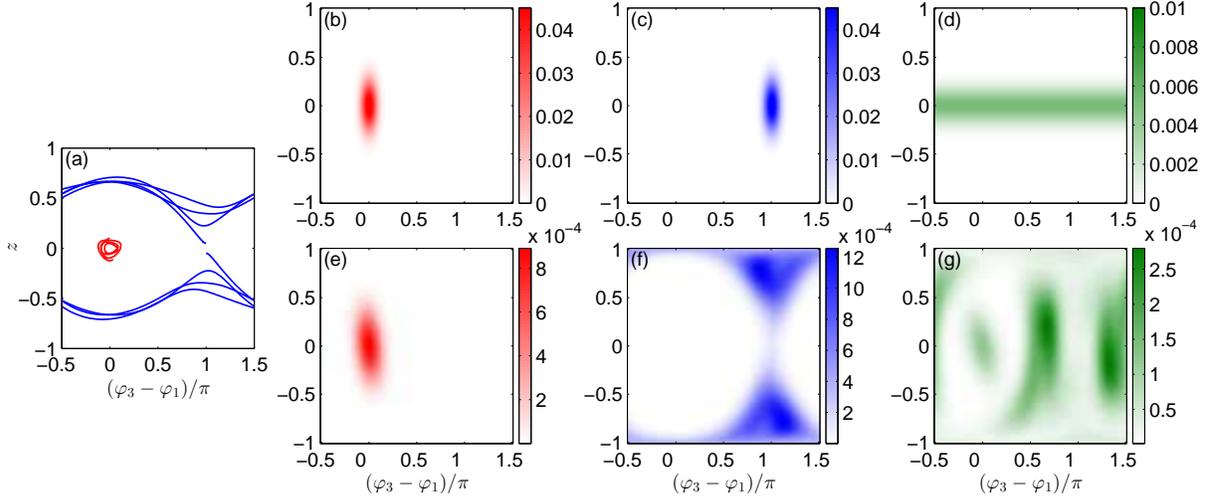}
\caption{\label{fig:husimi}
(Color online)
Semiclassical interpretation of breather state formation.
(a) Classical trajectories starting in the vicinity of the symmetric states
$(\alpha_1,\alpha_2,\alpha_3) = (1,0,1)/\sqrt{2}$ (red) and
the anti-symmetric states
$(1,0,-1)/\sqrt{2}$ (blue).
(b-g) The quantum dynamics of the Husimi $Q$ function follows
the classical phase space trajectories. 
(b,e) A BEC with a symmetric wave function $\ket{\Psi_+}$ 
remains approximately pure.
(c,f) A BEC with an anti-symmetric wave function$\ket{\Psi_-}$ is 
coherently split into two parts forming the breather state.
(d,g) A number state $\ket{\Psi_F}$ is also split into two parts, but
number fluctuations and correlations are less pronounced.
The Husimi function $Q(\alpha_1,\alpha_2,\alpha_3)$
is plotted as a function of the population imbalance 
$z = (|\alpha_3|^2 - |\alpha_1|^2)/n_{\rm tot}$ and the 
relative phase  $\phi_3 - \phi_1$ for $\alpha_2 = 0$ and 
$|\alpha_1|^2 + |\alpha_3|^2 = n_{\rm tot}$
at times $t=0$ (b-d) and $t=10 J^{-1}$ (e-g).
Here, $n_{\rm tot}$ denotes the total atom number at the 
respective time.
Parameters are $U = 0.1 J$, $\gamma_2 = 0.2 J$ and 
$N(0) = 60$. 
}
\end{figure*}

The formation of breather states can be understood to a large extent
within a semi-classical phase space picture. 
Any quantum state can be represented by a quasi distribution function
on the associated classical phase space without loss of information, such
as the Wigner or the Husimi function \cite{Gard04}. In the following, we
make use of both distribution functions which are defined as
\be
   Q(\alpha_1,\ldots,\alpha_M;t) := 
     \langle \alpha_1,\ldots,\alpha_M | \hat \rho(t) | 
            \alpha_1,\ldots,\alpha_M \rangle
\ee
and
\bea
   && \mathcal{W} := 
    \frac{1}{\pi^M} \int \prod \nolimits_j  d^2 \beta_j  \;
       \exp \left[ \sum \nolimits_j \alpha_j \beta_j^* - \alpha_j^* \beta_j \right]
           \nn \\
  && \quad \times \langle \alpha_1 - \beta_1 , \ldots , \alpha_M - \beta_M | 
       \hat \rho |   \alpha_1 + \beta_1 , \ldots , \alpha_M + \beta_M \rangle  
   \nn
\eea
respectively. Here, $|\alpha_j\rangle$ is a Glauber coherent state 
in the $j$th well and $M$ is the number of lattice sites. The evolution
equations of these distribution functions can be calculated systematically
using the operator correspondence discussed in \cite{Gard04}. The
evolution equation for the Wigner function is discussed in detail
in appendix \ref{sec:wigner}.

A general feature is that the dynamics of the phase space quasi distribution
functions is, to leading order in $1/N$, given by a classical Liouville equation,
\be
   \frac{\partial Q}{\partial t} = - \sum_j\left( \frac{\partial}{\partial \alpha_j} \dot \alpha_j
       + \frac{\partial}{\partial \alpha_j^*} \dot \alpha_j^* \right) Q
   + \mbox{noise}.
   \label{eqn:husimidyn}
\ee
Due to the structure of the evolution equation (\ref{eqn:husimidyn}), 
the `classical' Liouvillian flow provides the skeleton of the quantum 
dynamics (see M. Berry's quote in~\cite{Ber97}) of the Husimi or 
Wigner function, whereas the quantum corrections vanish with 
increasing particle number as $1/N$ \cite{07phase}. 
In particular, the Liouvillian approximation neglects phase-space 
interference effects as well as (anti-)diffusion terms which lead
to an elongation of the Wigner- and the 
Husimi-function \cite{07phaseappl}.
The associated classical flow is given by the dissipative discrete 
Gross-Pitaevskii equation (DGPE) \cite{Livi06,08mfdecay,Ng09}
\be
   i \dot \alpha_j = -J (\alpha_{j+1} - \alpha_{j-1}) + U |\alpha_j|^2 \alpha_j
      - i \gamma_j \alpha_j/2 \, .
     \label{gp}
\ee

Figure \ref{fig:husimi} (a) shows three trajectories of the DGPE for 
different initial values of the $\alpha_j = |\alpha_j| e^{i \phi_j}$. 
We have plotted
the evolution of the population imbalance between the first and third
site $z = (|\alpha_3|^2 - |\alpha_1|^2)/n_{\rm tot}$ vs. the relative phase  
$\Delta \phi = \phi_3 - \phi_1$. One observes that the trajectory 
starting at $\Delta \phi=0$ (red) is dynamically stable, such that 
it remains in the vicinity of the point $(z,\Delta \phi) = (0,0)$ for all times.
In contrast, trajectories starting close to $(z,\Delta \phi) = (0,\pi)$
converge to regions with either $z>0$ or $z <0$. These regions 
correspond to self-trapped states, which are known from the 
non-dissipative case \cite{Milb97,Smer97,Albi05}. For $\gamma_2 > 0$, these 
states become \emph{attractively stable}, which enables the 
dynamic formation of breather states.

The corresponding quantum dynamics of an initially pure BEC with 
a (anti-) symmetric $\ket{\Psi_{\pm}}$ wave function
is shown in Fig.~\ref{fig:husimi} (b,e) and (c,f), respectively. 
The Husimi function of the initial states are localized
around $(z,\Delta \phi) = (0,0)$ and  $(z,\Delta \phi) = (0,\pi)$  as shown in Fig.~\ref{fig:husimi} (b,c).
The DGPE then predicts the flow of the Husimi function on a coarse grained
scale. Trajectories starting in the vicinity of $(z,\Delta \phi) = (0,0)$ remain
close to their initial states and so does the Husimi function of the 
symmetric state $\ket{\Psi_+}$. In contrast, the Husimi function 
splits up into two fragments localized in the self-trapping regions
of phase space for the anti-symmetric initial state $\ket{\Psi_-}$ --
a breather state is formed.
Finally, in Fig.~\ref{fig:husimi} (d) the Husimi function of a Fock state is depicted. In
this case also the dynamics leads to the split of the function in two parts, as Fig.~\ref{fig:husimi} (g) illustrates.
However, the number fluctuations and correlations are less pronounced.

\begin{figure}[tb]
\centering
\includegraphics[width=7cm, angle=0]{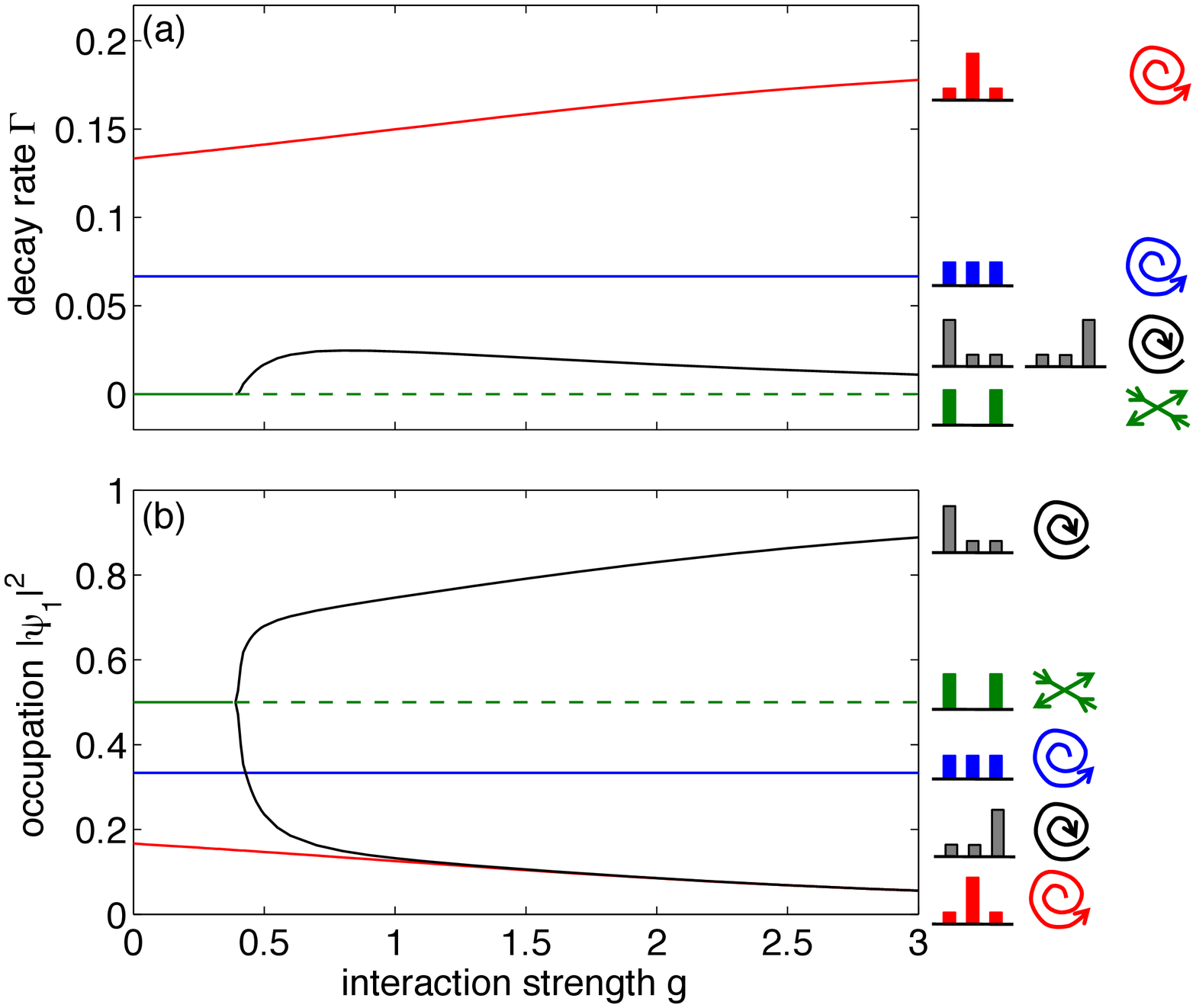}
\caption{\label{fig:trimer_mf2}
(Color online)
Properties of the meta-stable solutions of the non hermitian DGPE 
(\ref{eqn:dgpe-stat}) for $J=1$ and $\gamma_2 = 0.2$ as a function 
of the interaction strength $g=Un_{\rm tot}$. (a) Decay rate per atom $\Gamma$ 
and (b) relative occupation of the first well $|\alpha_1|^2$. The
icons on the right indicate the density distribution in the three wells 
and the dynamical stability for large $g$.
}
\end{figure}

The semi-classical picture predicts the fragmentation of the condensate
but, of course, cannot assert the coherence and thus the entanglement 
of the fragments which is a genuine quantum feature. However, it
correctly predicts the stability of an initial state and the emergence 
of breathers.
Thus we can infer the critical interaction strength for the transition 
to the breather regime from the associated ``classical'' dynamics. To this 
end we analyze the meta-stable states of the DGPE which are defined
as the solutions of the equation
\be
  - J (\alpha_{\ell-1} + \alpha_{\ell+1} ) + U |\alpha_{\ell}|^2 \alpha_{\ell}
  - i \frac{\gamma}{2} \delta_{\ell,2} \alpha_{\ell}
  = \left( \mu - i {\Gamma}/{2} \right) \alpha_{\ell}.
 \label{eqn:dgpe-stat} 
\ee
Here and in the following, we denote by 
$\vec \alpha = (\alpha_1,\ldots,\alpha_M) ^t$
the vector of all amplitudes $\alpha_j$. 
The meta-stable states are not stationary states of the 
DGPE in a strict sense, 
as the norm and thus the effective nonlinearity
$g = U \| \vec \alpha \|^2$ decays with a rate $\Gamma$. 
However, if decay is slow enough 
and if the solutions are dynamically stable, the time evolution will 
follow these quasi steady states adiabatically (cf., e.g., \cite{Schl06}).

The properties of the meta-stable states, their decay rate and 
their density distribution are summarized in Fig.~\ref{fig:trimer_mf2}
as a function of the effective nonlinearity $g$.
In the linear case $g=0$, three solution exist which are 
obtained by a simple diagonalization of the single-particle
Hamiltonian. Of particular interest is the anti-symmetric
state
\be
  \vec \alpha_{\rm as} = \frac{1}{\sqrt{2}} (1, 0, -1),  
  \label{eqn_initial-as} 
\ee
which exists for all $g$ and has a vanishing decay rate $\Gamma$.
With increasing interaction strength $g$, new solutions come into being. 
At a critical value $g_{\rm cr} =  0.4$, the anti-symmetric state 
$\vec \alpha_{\rm as}$ bifurcates and two breather solutions emerge. 
These breathers are strongly localized in one of the non-decaying 
wells $j=1,3$. Due to the symmetry of the system, both have the 
same decay rate $\Gamma$.

\begin{figure}[tb]
\centering
\includegraphics[width=6.5cm, angle=0]{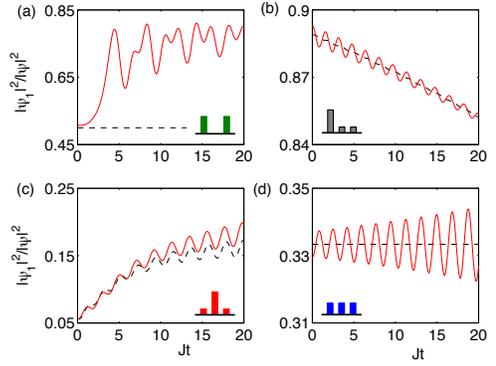}
\caption{\label{fig:trimer_stab}
(Color online)
Analysis of the dynamical stability of the meta-stable solutions of the 
DGPE (\ref{eqn:dgpe-stat}) for $J=1$, $Un_{\rm tot}=3$ and $\gamma_2 = 0.2$.
The dynamics has been simulated starting from a meta-stable state 
(black line) and the state plus a small perturbation of order $0.01$ 
(red line). Plotted is the relative occupation of the first well 
$|\alpha_1(t)|^2/\|\vec \alpha(t) \|^2$.
The density distribution of the initial states are 
illustrated by the icons in the corners: (a) the anti-symmetric state,
(b) a breather in the first well, (c) a breather in the leaky second well
and (d) the balanced state.
}
\end{figure}

For weak interactions, the state $\vec \alpha_{\rm as}$ dominates
the dynamics as its decay rate $\Gamma$ vanishes. However,
this is no longer possible  for $g > g_{\rm cr}$ as these states become
dynamically unstable as shown in Fig.~\ref{fig:trimer_stab} (a). 
Instead, the breathers dominate the dynamics. 
Their decay rate is rather small \cite{Ng09,Flac08,Camp04,Koh12} 
and, most importantly, they are 
attractively stable as shown in Fig.~\ref{fig:trimer_stab} (b). Thus, a 
breather is formed dynamically during the time evolution for most 
initial conditions if $g$ is large enough. 
The remaining meta-stable states are marginally unstable as 
shown in Fig.~\ref{fig:trimer_stab} (c,d).

\section{Decay in extended lattices}
\label{sec:lattice}

Next, we are going to discuss how localized single particle 
loss affects the dynamics in a more realistic extended lattice.
Also in this case a breather emerges when the 
interaction strength exceeds a critical value. In the following 
we will analyze the breather formation quantitatively 
and derive a formula for the critical interaction strength,
which depends on the size of the optical lattice.
The results presented in this section should be observable in
ongoing experiments with ultracold bosons in quasi
one dimensional optical lattices~\cite{Geri08,Wurt09,Guar11}.
As exact numerical simulations of the many-body quantum
dynamics are no longer possible for extended lattices with
many atoms, we use the truncated Wigner method 
(see appendix \ref{sec:wigner} for details). 
This approximate method is appropriate for a system with
large filling factors, since in this case the error induced from 
the truncation vanishes as $1/N$~\cite{07phase,07phaseappl}.
More importantly for our case, the truncated Wigner method 
can describe the deviation from a pure BEC state, in contrast 
to pure mean-field models~\cite{Ng09,Livi06,Hen12}. 

\begin{figure}[tb]
\centering
\includegraphics[width=8cm, angle=0]{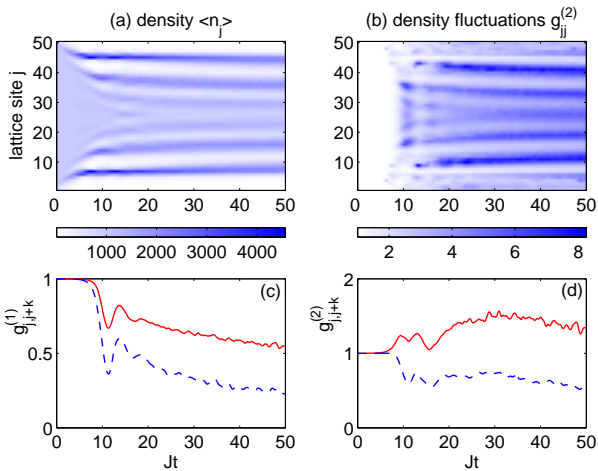}
\caption{\label{fig:lattice_dyn}
(Color online)
Dynamics of a leaky Bose-Hubbard chain with 50 wells. We have plotted
(a) the atomic density $\langle \hat n_j(t) \rangle$,
(b) the density fluctuations $g^{(2)}_{j,j}(t)$ in each lattice site
as well as (c) the phase coherence $g^{(1)}_{j,j+k}(t)$
and (d) the density-density correlations $g^{(2)}_{j,j+k}(t)$ 
between site $j=25$ and the neighboring sites 
$k=1$ (solid red line), $k=2$ (dashed blue line).
Parameters are $UN(0)=25 J$, $\gamma_1=2 J $, 
$M=50$ and $\rho(t=0)=N/M=1000$.
}
\end{figure}

\subsection{Breather state formation}
In the following we consider an extended optical lattice consisting 
of $M=50$ sites with periodic boundary conditions unless states 
otherwise. Loss occurs from the lattice site $j=1$ only. 
As an initial state we assume a pure BEC which is moved at constant 
speed \cite{Sias07} or accelerated \cite{Peik97} to the edge of the 
first Brillouin zone. Hence the quantum state of the atoms at $t=0$ 
is given by 
\be
    |\Psi(0) \rangle  = \frac{1}{\sqrt{N!}} 
            (\sum \nolimits_j \psi_j \hat a_j^\dagger)^N \ket{0}
\ee 
with $\psi_j = (-1)^j/\sqrt{M}$, which generalizes the antisymmetric initial state 
$\ket{\Psi_-}$ discussed for the triple-well trap.  We consider the case of 
large filling factors, with $N/M=1000$ in all simulations.

For weak interactions the quantum state remains close to a pure 
BEC during the decay, such that all coherence functions are 
approximately one. The dynamics changes dramatically for 
strong interactions as shown in Fig.~\ref{fig:lattice_dyn}. 
The phase coherence $g_{j,k}^{(1)}$ between adjacent wells 
is lost after a short transient period, indicating the dynamical 
instability of the condensate.
At the same time the number fluctuations $g_{j,j}^{(2)}$ rapidly
increase as shown in Fig.~\ref{fig:lattice_dyn} (b). This reveals
a strong spatial bunching of the atoms as expected for a breather
state. This feature of the dissipative equilibrium state is in strong 
contrast to the non-dissipative case, where repulsive interactions
\emph{suppress} number fluctuations in thermal equilibrium.
Part (d) of the figure reveals the second characteristic trait
of the breather state. Strong anti-correlations with 
$g_{j,j+2}^{(2)} \approx 0.5$ are observed between the 
site $j=25$ and the next-to-nearest neighbor. 
No anti-correlations are observed for the direct neighbor, 
as breathers can extend over more than one site in an extended lattice. 
We thus conclude that the atoms tend to bunch at one site of the 
lattice, leaving the neighboring sites essentially empty. This is exactly 
the signature of the breather state in the extended lattice, which we 
have discussed above for the trimer case. The position of the individual
breathers in this breather state is random due to the quantum
fluctuations. We note that it can be experimentally
easier to prepare breather states starting from
a Mott insulator instead of a BEC at the band
edge. Simulations for small lattices show the de-
velopment of strong density anti-correlations and
multi-particle entanglement also in this case.

\begin{figure}[tb]
\centering
\includegraphics[width=8.5cm, angle=0]{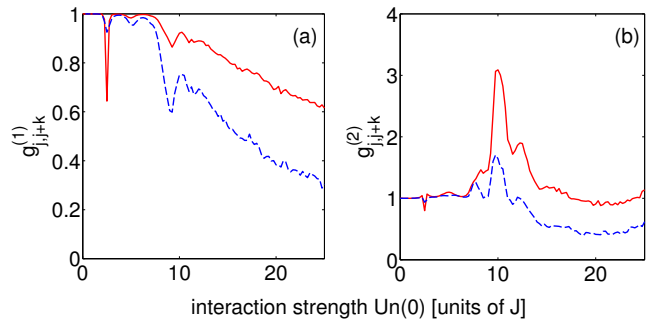}
\caption{\label{fig:lattice_vsu}
(Color online)
Transition to the breather regime in an open optical lattice. 
Shown is (a) the the phase coherence $g^{(1)}_{j,j+k}$
and (b) the number correlation function $g^{(2)}_{j,j+k}$
between site $j=9$ and the neighboring sites  
$k=1$ (solid red line) and $k=2$ (dashed blue line), respectively,
after a fixed propagation time $t_{\rm final} = 50 J^{-1}$.
One observes a sharp transition when the interaction strength 
$UN(0)$ exceeds a critical value of approx. 
$2.5J$.
Parameters are $\gamma_1=2 J $, $M=50$ 
and the atomic
density is $\rho(t=0) = N/M=1000$.
}
\end{figure}

The transition to the breather regime for strong interactions is further
analyzed in Fig.~\ref{fig:lattice_vsu}, which shows the first and second
order coherence functions as a function of the interaction strength for 
a fixed propagation time $t_{\rm final}=50\, J^{-1}$ at the reference 
site $j=9$. For $UN(0) \lesssim 2.5\, J$ the phase coherence 
between neighboring sites is preserved, while the atoms decay from the lattice. 
For stronger interactions, however, phase coherence is lost and the BEC fragments into a breather state.
The second order correlation function $g_{j,j+2}^{(2)}$ reveals the 
existence of strong anti-correlations for large values of $U$. However,
we observe $g_{j,j+2}^{(2)}> 1$ in the vicinity of the transition point.
This is a consequence of the localization of the breathers, which 
becomes tighter with increasing $U$ \cite{Flac08}. Directly above the 
transition breathers exist, but typically extend over several 
lattice sites, such that we observe positive correlations at this length scale.
Moreover, the formation of breathers suppresses the decay from 
the lattice, that is, the total particle number decreases more slowly.
This is due to the strong localization of the breathers preventing atoms
from tunneling to the leaky lattice sites.

Note that the coherence functions show the same qualitative 
behaviour if another lattice site is chosen as a reference site instead
of $j=25$ or $j=9$.
The oscillations we observe in Fig.~\ref{fig:lattice_vsu} (b) and for
intermediate values of $U$ are just a manifestation of the temporal 
oscillations of the $g^{(1)}$ and $g^{(2)}$, as shown in 
Fig.~\ref{fig:lattice_dyn} (c) and (d). 

\subsection{Critical interaction strength}

\begin{figure}[tb]
\centering
\includegraphics[width=8cm, angle=0]{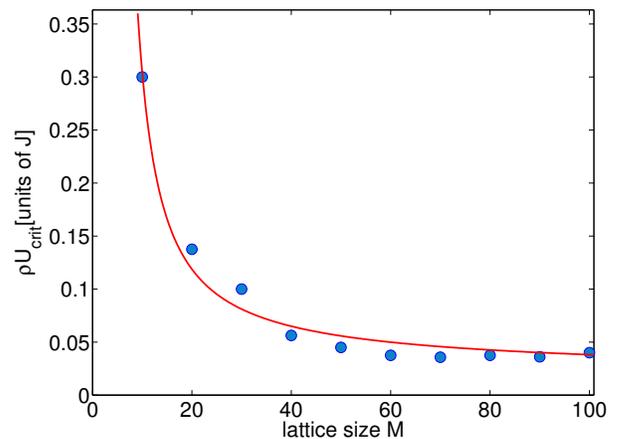}
\caption{\label{fig:LvsM}
The critical nonlinearity $\rho U_{\rm crit}$, at which breathers start to form, as a function of the lattice size $M$. 
Numerical results  using the truncated Wigner method (blue circles) are compared to a fit using equation (\ref{eqn:Ucrit}) (red line).
The fitting parameters we found are $L = 0.075$ with bounds $(0.054,0.096)$ and $P_{\rm th} = 1 - 2.4 \times 10^{-7}$ with bounds 
$(1 + 4.54 \times 10^{-7},1 - 9.3 \times 10^{-7})$, while the summed square of residuals is $SSE=1.2 \times 10^{-3}$. The other parameters are $\gamma_1=2J$ and 
$\rho(t=0) = N/M=1000$.
}
\end{figure}

Breather formation sets in abruptly when the interaction strength exceeds a critical value $U_{\rm crit}$. 
Extensive numerical simulations show that the transition point depends on the size of the lattice, i.e. the number of 
sites $M$, as shown in Fig.~\ref{fig:LvsM}. As the lattice becomes larger, breather formation is facilitated such that the 
critical value $U_{\rm crit}$ decreases rather rapidly. In these simulations, $U_{\rm crit}$ was determined as follows.
{After a fixed propagation time we find the values of the density fluctuations $g^{(2)}_{j,j}$ for different interaction strengths $U$
and for various lattice sites $j$. 
We identified the critical interaction as the maximum
interaction strength in which the density fluctuations at all sites $j$ differ from the value in the non-interacting case, 
$g^{(2)}_{j,j}=1$, by less than $5\%$.
 In all simulations we have used $\gamma_1=2J$ and the same initial density, $\rho(t=0) = N/M=1000$.}
In the following we derive a formula for the critical interaction strength, which will also clarify  the microscopic origin of 
breather formation and its connection to the self-trapping effect. Our considerations follows the reasoning presented in \cite{Hen12} 
for the analogous mean-field system.

As shown in~\cite{Hen12}, breathers formation is a local process, which occurs if the local effective nonlinearity exceeds a critical value $L$
\be
    U n_j / J  \le L
   \label{eqn:lcrit}
\ee
for at least one lattice site $j$. Then the nonlinearity is strong enough 
to induce self-trapping at the respective lattice site (cf.~also \cite{Albi05,Milb97,Smer97}). Starting from this local ansatz, 
the critical interaction strength can be inferred as follows. Breathers are observed if the probability to satisfy condition 
(\ref{eqn:lcrit}) exceeds a certain threshold value
\be
    {\rm prob}(\exists~j: n_j > JL/U) \ge P_{\rm th}.
  \label{eqn:pth}
\ee
Hennig and Fleischmann~\cite{Hen12} furthermore argue that the probability to observe a certain atom number $n_j$ follows a 
Poissonian distribution in the diffusive regime, such that the cumulative distribution function is given by 
${\rm prob}(n_j < n_{\rm crit}) = 1 - e^{-M n_{\rm crit}/N}$. Using this result for a single lattice site, one calculate the probability
to find at least one $n_j \ge n_{\rm crit} = JL/U$:
\bea
   {\rm prob}(\exists~j: n_j > n_{\rm crit}) &=& 
      1 -  {\rm prob}(n_j < n_{\rm crit} \, \forall j)   \nn \\
   &=& 1- (1- e^{-M n _{\rm crit}/N})^M. \nn
\eea
Substituting this result into equation (\ref{eqn:pth}) and solving for $U$ then yields the following condition for the onset of breather formation
\be
   U \rho \ge U_{\rm crit} \rho =  \frac{-JL}{
                     \ln [1 - (1-P_{\rm th})^{1/M}]} \, ,
  \label{eqn:Ucrit}
\ee
where $\rho = N/M$ is the atomic density.

The analytic prediction (\ref{eqn:Ucrit}) depends on two parameters
$L$ and $P_{\rm th}$, which are used as fit parameters to model the numeric results. This fit yields an excellent 
agreement with the numeric results as shown in Fig.~\ref{fig:LvsM}. We stress that the decrease of $U_{\rm crit}$ 
with increasing lattice site cannot be modeled by a simple algebraic or exponential decay. Notably, we obtain 
significantly smaller values for $U_{\rm crit}$ than in \cite{Hen12}. This is attributed to the fact that the unstable
initial state considered in this paper, a BEC at the band edge, has a higher energy and thus fragments into breathers much more easily.

\section{Conclusions and outlook}

Recently, there has been a great interest in engineering quantum dynamics
by using dissipation in various systems~\cite{Diel08,Diel11,Vers09,Kast11,Krau10,Tom12,Reg11,Geri08,Wurt09,Baron13,11leaky1,11leaky2,Kepe12,Angl97,08stores,09srlong,Ng09,Afek10,Livi06,08mfdecay,Guar11,Hen12}. 
In this paper, we report the effects of an elementary dissipation mechanism, the localized single particle loss,
in a BEC loaded in a deep optical lattice. Particle losses combined with strong interparticle interactions
and discrete geometry can deterministically lead to the formation of quantum superpositions of discrete breathers.
For a small trimer system we have discussed the properties of these ``breather states'' in detail including entanglement, decoherence and
possible applications in precision quantum interferometry. A semiclassical interpretation of breather state formation
has revealed the connection to a classical bifurcation of the associated mean-field dynamics.
Furthermore, we have studied the dynamical formation of breather states in extended lattices and we have derived a
formula which predicts the critical interaction strength, in which the breathers start to form, in lattices with different size.
The formation and the properties of these structures could be readily observed in ongoing experiments with ultracold atoms
in optical lattices~\cite{Geri08,Wurt09,Bakr09,Sher10,Baron13,Albi05}.

Nonlinear structures, like bright or dark solitons,
are well known in the context of the Gross-Pitaevskii equation
(see for example~\cite{Abd11,Alex12,Achi12}). However, this equation
cannot give us any information about the quantum nature of the problem, these structures are ``classical'' objects.
With the present work we open a new direction: stable nonlinear structures that exhibit purely quantum properties,
like entanglement. These properties cannot be studied anymore with a simple
GP equation (or DGPE for discrete systems) and one should go beyond them to support
state-of-the-art experiments.

\begin{acknowledgments}

We acknowledge financial support by the Max Planck Society and 
the Deutsche Forschungsgemeinschaft (DFG) via the Forschergruppe 
760 (grant number WI 3426/3-1) and the Heidelberg 
Graduate School of Fundamental Physics (grant number GSC 129/1).

\end{acknowledgments}

\appendix

\section{Entanglement criterion}
\label{sec:ent}

In this section we provide a detailed derivation of the 
entanglement criterion based on (\ref{eqn:ent_para})
which is adapted to the NOON states discussed in the present paper.
This result generalizes established entanglement
criteria in terms of spin squeezing \cite{Sore01} and is derived
in a similar way.
In contrast to spin squeezing inequalities, it shows that
a state is entangled if the variance defined below in (\ref{eqn:ent_para2}) is
\emph{larger} than a certain threshold value.

We assume that the many-body quantum state $\hat \rho$ is 
decomposed into a mixture of pure states
\bea
   \hat \rho &=& \sum_a p_a \hat \rho_a \nn \\
    &=&  \sum_a p_a \ket{\psi_a}\bra{\psi_a},
   \label{eqn:rhodecompose}
\eea
where every pure state $\hat \rho_a = \ket{\psi_a}\bra{\psi_a}$ 
has a fixed particle number $N_a$. Note that the quantum jump 
simulation of the dynamics directly provides such a decomposition.
We define the entanglement parameter
\bea
       \label{eqn:ent_para2}    
    E_{r,q} &:=& \langle (\hat n_r - \hat n_q)^2 \rangle
     - \langle \hat n_r - \hat n_q \rangle^2 
     - \langle \hat n_r + \hat n_q \rangle  \\
    &&  - \frac{1}{2} \sum_{a,b} p_a p_b 
     \left[ \langle (\hat n_r - \hat n_q) \rangle_a 
         - \langle (\hat n_r - \hat n_q) \rangle_b \right]^2 \nn
\eea
for the sites $r$ and $q$. In this expression $\langle \cdot \rangle_{a,b}$ 
denotes the expectation value in the pure state $\ket{\psi_{a,b}}$. 
Now we can proof that $E_{r,q}<0$ for every separable state such 
that a value $E_{r,q}>0$ unambiguously reveals the presence of
many-particle entanglement.
Note that $E_{r,q}$ provides and entanglement criterion,
it is not a quantitative entanglement measure in the strict sense.

To proof this statement we consider an arbitrary separable
state and show that $E_{r,q}<0$ for this class of states. 
If a pure state $\hat \rho_a$ is separable, it can be written 
as a tensor product of single particle states
\be 
   \hat \rho_a =   \hat \rho_a^{(1)} \otimes   \hat \rho_a^{(2)} 
           \otimes \cdots \otimes   \hat \rho_a^{(N_a)},
   \label{eqn:psstate}
\ee
We furthermore introduce the abbreviation
\be
  \hat S_\pm := \hat n_r \pm \hat n_q.
\ee
This operator is also written as a symmetrized
tensor product of single-particle operators
\bea
   \hat S_\pm = \sum_{k=1}^{N_a} \eye \otimes \cdots \otimes \eye 
       \otimes \hat s_\pm^{(k)} \otimes \eye \otimes \cdots \otimes \eye,
\eea
where the superscript $(k)$ denotes that the single-particle operator $\hat s_\pm^{(k)}$ 
acts on the $k$th atom. The single-particle operators are given by
\be
    \hat s_\pm = \ket{r}\bra{r} \pm  \ket{q}\bra{q},
\ee
where $\ket{r}$ is the quantum state where the particle is 
localized in site $r$.

For a separable pure state  $\hat \rho_a$, the expectation 
values of the population imbalance 
$ \langle \hat S_- \rangle_a = \tr [\hat \rho_a \hat S_-]$
and its square can be expressed as
(dropping the subscript $a$ for notational clarity)
\bea
   \langle \hat S_- \rangle &=& \sum_{k=1}^N \tr\left[ \rho^{(k)} \hat s_-^{(k)}  \right] \nn \\
   \langle \hat S_-^2 \rangle &=& \sum_{j \neq k}^N \tr \left[ (\rho^{(j)} \otimes \rho^{(k)} )
            (\hat s_-^{(j)} \otimes \hat s_-^{(k)}) \right]  \nn \\
    &&  + \sum_{j=1}^N  \tr\left[ \rho^{(j)} \hat s_-^{(j)2}  \right] \nn \\
     &=& \sum_{j,k=1}^N \tr\left[ \rho^{(j)} \hat s_-^{(j)}  \right] \tr\left[ \rho^{(k)} \hat s_-^{(k)}  \right] \nn \\
        && \!\!\!\!\! -  \sum_{j=1}^N \tr\left[ \rho^{(j)} \hat s_-^{(j)}  \right] \tr\left[ \rho^{(j)} \hat s_-^{(j)}  \right] 
           + \sum_{j=1}^N \tr\left[ \rho^{(j)} \hat s_-^{(j)2}  \right] \nn \\
      &=& \langle \hat S_- \rangle^2 +
           \sum_{j=1}^N  \tr\left[ \rho^{(j)} \hat s_-^{(j)2}  \right]   - \left\{ \tr\left[ \rho^{(j)} \hat s_-^{(j)}  \right] \right\}^2. \nn
\eea
Using $\tr[ \rho^{(j)} \hat s_-^{(j)2} ] = \tr[ \rho^{(j)} \hat s_+^{(j)}]$ we thus find that 
every pure products state $\hat \rho_a$ satisfies the condition
\be
   \langle \hat S_-^2 \rangle_a - \langle \hat S_- \rangle^2_a \le \langle \hat S_+ \rangle_a \, .
\ee

If the total quantum state $\hat \rho$ is separable, such that it can be written as a mixture of
separable pure states (\ref{eqn:rhodecompose}), the expectation values are given by
\bea
   && \langle \hat S_-^2 \rangle = \sum_a p_a \langle \hat S_-^2 \rangle_a  \\
   && \qquad \le \langle \hat S_+ \rangle + \sum_a p_a  \langle \hat S_- \rangle_a^2 \nn \\
   && \langle \hat S_- \rangle^2 = \sum_{a,b} p_a p_b \langle \hat S_- \rangle_a \langle \hat S_- \rangle_b \\
   && \qquad = \sum_a p_a  \langle \hat S_- \rangle_a^2
            - \frac{1}{2} \sum_{a,b} p_a p_b \left[ \langle \hat S_- \rangle_a - \langle \hat S_- \rangle_b \right]^2. \nn
\eea
We thus find that every separable quantum state satisfies the following inequality
for the variance  of the population imbalance $\hat S_-$: 
\be
   \langle \hat S_-^2 \rangle - \langle \hat S_- \rangle^2
   \le \langle \hat S_+ \rangle +
    \frac{1}{2} \sum_{a,b} p_a p_b \left[ \langle \hat S_- \rangle_a - \langle \hat S_- \rangle_b \right]^2.
\ee
This inequality for separable quantum states can be rewritten as
\be
    E_{r,q} < 0
\ee
in terms of the entanglement parameter (\ref{eqn:ent_para2}).

\section{Truncated Wigner function dynamics}
\label{sec:wigner}

In this appendix, we will explicitly derive the evolution equation for the 
Wigner function which corresponds 
to the master equation (\ref{eqn:master}).
To this end we use the following operator 
correspondences~\cite{Gard04}:
\bea
   \hat{a}_{j}\hat{\rho} &\leftrightarrow& 
      \left(\alpha_{j}+\frac{1}{2}\frac{\partial}{\partial\alpha_{j}^*}\right)\mathcal{W}, \\
   \hat{\rho}\hat{a}_{j} &\leftrightarrow &
       \left(\alpha_{j}-\frac{1}{2}\frac{\partial}{\partial\alpha_{j}^*}\right)\mathcal{W}, \\
   \hat{a}_{j}^\dagger\hat{\rho} &\leftrightarrow &
      \left(\alpha_{j}^*-\frac{1}{2}\frac{\partial}{\partial\alpha_{j}}\right)\mathcal{W}, \\
  \hat{\rho}\hat{a}_{j}^\dagger &\leftrightarrow &
     \left(\alpha_{j}^*+\frac{1}{2}\frac{\partial}{\partial\alpha_{j}}\right)\mathcal{W},
\eea
where $\alpha_j$ are the eigenvalues of the destruction operator:
\be
   \hat{a}_j|\alpha_j\rangle = \alpha_j |\alpha_j\rangle, \phantom{0} 
       \langle\alpha_j|\hat{a}_j^\dag =\alpha^*_j \langle\alpha_j|.
\ee
Substituting these correspondences in the master equation 
(\ref{eqn:master}), we obtain the 
following evolution equation for the Wigner function:
\begin{widetext}
\bea 
  \partial_t\mathcal{W} &=& 2J\sum_{j=1}^{M-1}\Im \bigg[
    \left(\alpha_{j}-\frac{1}{2}\frac{\partial}{\partial\alpha_{j}^*}\right)
    \left(\alpha_{j+1}^*+\frac{1}{2}\frac{\partial}{\partial\alpha_{j+1}}\right)
    - \left(\alpha^*_{j+1}-\frac{1}{2}\frac{\partial}{\partial\alpha_{j+1}}\right)
    \left(\alpha_{j}+\frac{1}{2}\frac{\partial}{\partial\alpha_{j}^*}\right)\bigg]\mathcal{W} 
    \nn \\
   &&  +U \sum_{j=1}^{M} \Im\left(\alpha_{j}-\frac{1}{2}\frac{\partial}{\partial\alpha_{j}^*}\right)^2
         \left(\alpha_{j}^*+\frac{1}{2}\frac{\partial}{\partial\alpha_{j}}\right)^2\mathcal{W}
         - \sum_{j=1}^{M}\frac{\gamma_j}{2}\bigg[  
              \left(\alpha^*_{j}-\frac{1}{2}\frac{\partial}{\partial\alpha_{j}}\right)
              \left(\alpha_{j}+\frac{1}{2}\frac{\partial}{\partial\alpha_{j}^*}\right)  
            \nn  \\
   && +\left(\alpha_{j}-\frac{1}{2}\frac{\partial}{\partial\alpha_{j}^*}\right)
           \left(\alpha_{j}^*+\frac{1}{2}\frac{\partial}{\partial\alpha_{j}}\right) 
            -2\left(\alpha_{j}+\frac{1}{2}\frac{\partial}{\partial\alpha_{j}^*}\right)
           \left(\alpha_{j}^*+\frac{1}{2}\frac{\partial}{\partial\alpha_{j}}\right)\bigg]\mathcal{W}.
\label{eqn-wig}
\eea
As one can easily see the above equation includes not only first and second order derivatives, 
but also third order ones arising from the interaction term (the $U$-dependent term in the second line of equation (\ref{eqn-wig})).
These third order derivatives makes the equation
quickly unstable, so an approximate method is needed. One technique that 
is widely used in optical systems is the \emph{truncated Wigner} method~\cite{Sina02,Wern97},
which is a good approximation as far as the mode occupation numbers are large.
In this approximation one neglects all the terms that include third order derivatives, thus we have the equation
\bea
   \label{epn-fok}
   \partial_t\mathcal{W} &=& \sum_j\frac{\partial}{\partial x_j}\Bigg[J(y_{j+1}+y_{j-1}) 
    U(y_j - x_j^2y_j - y_j^3) + \frac{\gamma_j}{2}x_j\Bigg] \mathcal{W}  \\
    && + \sum_j\frac{\partial}{\partial y_j}\bigg[-J(x_{j+1} +x_{j-1}) - U(x_j - x_j y_j^2 - x_j^3) 
       + \frac{\gamma_j}{2}y_j \bigg]  \mathcal{W} 
    + \frac{1}{2} \sum_j \frac{\gamma_j}{4} \left(\frac{\partial^2}{\partial x_j^2} 
       + \frac{\partial^2}{\partial y_j^2}\right)\mathcal{W}, \nn
\eea
where $x_j,y_j$ are the real and imaginary part of $\alpha_j$ respectively.
\end{widetext}

Equation (\ref{epn-fok}) is a Fokker-Planck equation, thus it can be rewritten in the language of stochastic differential or Langevin equations. 
To be more precise, consider the Fokker-Planck equation of the form~\cite{Gard09}:
\bea 
 \label{eqn-stoch}
    \partial_t\mathcal{W} &=& - \sum_j \frac{\partial}{\partial z_j} A_j(\textbf{z},t)\mathcal{W}  \\
   && +\frac{1}{2}\sum_{j,k}\frac{\partial}{\partial z_j}\frac{\partial}{\partial z_k}
         \left[\textbf{B}(\textbf{z},t)\textbf{B}^T(\textbf{z},t)\right]_{jk}\mathcal{W}, \nn 
\eea
where the diffusion matrix $D=\textbf{B}\textbf{B}^T$ is positive definite. 
Now, we can write equation (\ref{eqn-stoch}) as a system of stochastic equations:
\be
\frac{d\textbf{z}}{dt}=A(\textbf{z},t)+\textbf{B}(\textbf{z},t)\textbf{E}(t),
\ee
where the real noise sources $E_j(t)$ have zero mean and satisfy 
$\langle E_j(t) E_k(t')\rangle=\delta_{jk}\delta(t-t')$. 
In our case, equation (\ref{epn-fok}) can be rewritten:
\bea
   && \frac{dx_j}{dt}=-J(y_{j+1}+y_{j-1}) - U(y_j - x_j^2y_j - y_j^3)   \nn \\
    && \qquad \qquad  -\frac{\gamma_j}{2}x_j + \frac{\sqrt{\gamma_j}}{2}\xi_j(t), \label{stoch1} \\
   && \frac{dy_j}{dt}=J(x_{j+1}+x_{j-1}) + U(x_j - x_j y_j^2 - x_j^3) \nn \\ 
     && \qquad \qquad   -\frac{\gamma_j}{2}y_j + \frac{\sqrt{\gamma_j}}{2}\eta_j(t), \label{stoch2}
\eea
where $\xi_j(t),\eta_j(t)$ for $j=1,...,M$ are $\delta$-correlated in time with zero mean. 
Here it must be noted that $\xi_j(t),\eta_j(t)$ are not real noise sources, but are included 
only to recapture the commutation relations of the operators.

\begin{figure}[tb]
\centering
\includegraphics[width=8cm, angle=0]{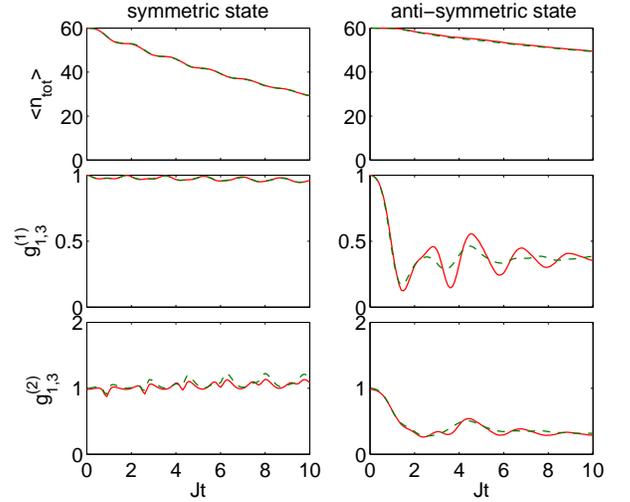}
\caption{\label{fig:comp}
A comparison of a truncated Wigner simulation (dashed green lines)
and a quantum jump simulation (solid red lines) shows a very good 
agreement.
Shown is the time evolution of 
the total atom number $\langle \hat n_{\rm tot} \rangle$,
the phase coherence $g^{(1)}_{1,3}$,
and the density-density correlations $g^{(2)}_{1,3}$.
The parameters and initial states are the same as in 
Fig.~\ref{fig:dynamics} with $U=0.1J$.
}
\end{figure}

As an initial state one uses a product state of the form
\bea
  |\Psi(t=0)\rangle = |\psi_1\rangle|\psi_2\rangle...|\psi_M\rangle,
  \label{init1}
\eea
where $|\psi_j\rangle$ is a Glauber coherent state in the $j$th well. This state represents a pure BEC in a grand-canonical framework.
The Wigner function of a Glauber coherent state $|\psi_j\rangle$ is a Gaussian,
\be
  \mathcal{W}(\alpha_j,\alpha_j^*)=\frac{2}{\pi}\exp\{-|\alpha_j-\psi_j|^2\} \, .
  \label{init2}
\ee
Thus we can take the initial values for $\alpha_j=x_j+iy_j$ to be Gaussian 
random numbers with mean $\psi_j$.
For a BEC in a Bloch state with quasi momentum $k$, we have
\be
  \psi_j=e^{ikj}\sqrt{\frac{N}{M}} \, .
  \label{init3}
\ee
In the text we consider a pure BEC accelerated to the 
edge of the Brillouin zone such that $k=\pi$.

The truncated Wigner method is used to calculate 
the evolution of expectation of symmetrized observables as follows. 
The Wigner function is treated  as a probability distribution 
in phase space. An ensemble oftrajectories is sampled
according to the Wigner function of the initial state and
the propagated according to Eqs.~(\ref{stoch1}) and (\ref{stoch2}).
Then one takes the stochastic average over this ensemble:,
\bea
  \langle O_j...O_k\rangle_{\rm sym} &=& \int \prod_{i=1}^M d^2 \alpha_i \, O_j...O_k \, \mathcal{W}(\alpha_1,\alpha_1^*,...) \nn \\
   &=&\frac{1}{N_T}\sum_{\ell=1}^{N_T}O_j...O_k
\eea
where $O_j$ stands for $\alpha_j$ or $\alpha_j^*$, $N_T$ is the number 
of trajectories and the subscript sym reminds us that only expectations values of
symmetrized observables can be calculated.

In Fig. \ref{fig:comp} we compare the results of the truncated Wigner 
approximation with the results of the exact quantum jump method
for the triple-well trap studied in Sec.~\ref{sec:trimer}.
The simulations show a very good agreement also in the regime of
strong interactions. The only small discrepancy is that oscillations 
of the correlation functions are slightly less pronounced.
As the truncated phase space approximations become more
accurate with increasing filling factors~\cite{07phase,07phaseappl}, 
we expect that the truncated Wigner simulations discussed in
Sec.~\ref{sec:lattice} are reliable both qualitatively and quantitatively.


\bibliographystyle{apsrev}

\end{document}